\documentclass[a4paper,11pt,nofootinbib,nobibnotes]{article}
\usepackage{jcappub}
\pdfoutput=1

\usepackage[T1]{fontenc}
\usepackage[utf8]{inputenc}
\usepackage{lmodern}
\usepackage{graphicx}
\usepackage{amsmath,amssymb,amsfonts}
\usepackage{hyperref}
\usepackage{physics}
\usepackage{braket}
\usepackage{overpic}
\usepackage[dvipsnames]{xcolor}

\setcitestyle{numbers,square}

\hypersetup{
    colorlinks=true,
    linkcolor=blue,
    citecolor=blue,
    urlcolor=blue
}

\graphicspath{{images/}}

\title{Adiabatic renormalization for modified dispersion relations in cosmology}

\author[a]{Christian Durán-Romero}
\author[a]{Luis J. Garay}
\author[a]{Mercedes Martín-Benito}
\author[b]{Rita B. Neves}

\affiliation[a]{Departamento de Física Teórica and IPARCOS,  
Universidad Complutense de Madrid, 28040 Madrid, Spain}
\affiliation[b]{School of Mathematical and Physical Sciences,  
University of Sheffield, Hicks Building, Hounsfield Road,  
Sheffield S3 7RH, United Kingdom}

\emailAdd{chduran@ucm.es}
\emailAdd{luisj.garay@ucm.es}
\emailAdd{m.martin.benito@ucm.es}
\emailAdd{rita.neves@sheffield.ac.uk}

\abstract{We investigate the behavior of scalar quantum fields in cosmological backgrounds under modified dispersion relations, specifically focusing on how ultraviolet asymptotics influence field quantization. We establish the conditions for both the validity of the adiabatic approximation and the unitary equivalence between quantizations defined via different time variables. Our analysis reveals that while superluminal modified dispersion relations consistently yield unitarily equivalent quantizations, asymptotically subluminal behaviors can lead to inequivalent physical descriptions. By applying adiabatic regularization to the two-point correlation function, we demonstrate that the ultraviolet scaling of the frequency uniquely dictates the required subtraction order. These results are illustrated through applications to standard, superluminal Corley--Jacobson, and Unruh dispersion relations.}

\begin{document}
\maketitle
\flushbottom
\section{Introduction}
\label{Section I}
The theory of cosmological inflation has become the prevailing paradigm in early universe cosmology. Initially proposed to resolve the flatness, horizon, and monopole problems of the standard Big Bang model~\cite{kolbturner}, inflation posits a brief epoch of accelerated expansion that drives the spatial curvature to near zero and stretches microscopic regions to cosmological scales. Beyond solving these problems, inflation provides a mechanism for the generation of primordial density perturbations: quantum fluctuations of the inflaton field are redshifted beyond the Hubble radius and later re-enter as classical seeds for large scale structure~\cite{mukhanov2005physical}. Observations of the Cosmic Microwave Background (CMB) anisotropies have confirmed the resulting nearly scale-invariant primordial power spectrum with remarkable precision~\cite{wmap2011,achúcarro2022inflationtheoryobservations}.

Notwithstanding these triumphs, a conceptual subtlety arises when one traces observable perturbation modes back to the onset of inflation. If inflation endures for more than the minimal sixty  $e$-foldings or so, then the physical wavelengths corresponding to CMB scales would have been shorter than the Planck length at the very beginning of the inflationary phase. At such trans-Planckian scales, the classical concept of spacetime and the standard methods of quantum field theory in curved backgrounds might lose their validity. This is the so-called trans-Planckian problem of inflationary cosmology~\cite{Brandenberger_2013}. One must confront the possibility that predictions based on extrapolating low-energy effective field theory may receive significant corrections from unknown ultraviolet (UV) physics, potentially altering the power spectrum.

Most trans-Planckian studies in cosmology assume that at the onset of inflation perturbations were in an adiabatic vacuum state, neglecting any preceding dynamics. Yet if new physics is operative at the Planck scale, its imprint may be even stronger during a pre-inflationary epoch~\cite{PhysRevD.63.123501}. This consideration opens a new window: rather than undermining the predictive power of inflation, the trans-Planckian problem may provide an observational handle on quantum gravity phenomena through subtle imprints in the primordial spectrum and higher-order correlators~\cite{Brandenberger2002}.

A closely related issue was identified in black hole physics. Hawking radiation is derived by tracing outgoing modes backward in time to arbitrarily short distances near the event horizon, where quantum gravitational effects are expected to dominate. Remarkably, analyses incorporating modified dispersion relations (MDRs)—in which the usual linear relation between frequency and wavenumber is deformed at high energies—have shown that the thermal character of Hawking radiation is robust over a wide range of UV deformations~\cite{corley1996, unruh1995sonic}. This suggests that MDRs can serve as useful phenomenological proxies for the unknown Planck-scale theory, encoding potential quantum gravity corrections in a simple, controllable manner.

In the inflationary context, MDRs play an analogous role. By substituting the standard dispersion relation with one that deviates at a characteristic UV scale, one aims to effectively model the influence of quantum gravity on the evolution of perturbation modes~\cite{Niemeyer_2001, Brandenberger2001}. Two archetypal classes of MDRs are frequently considered in the literature. The first one is given by the Corley--Jacobson (CJ) family \cite{corley1996}, which introduces smooth high-momentum corrections and can be further classified into superluminal or subluminal cases, depending on whether the dispersion relation remains monotonic or not in the UV regime. The second archetypal example is provided by the Unruh dispersion relation \cite{unruh1995sonic}, which imposes a frequency cutoff reflecting a maximal propagation speed in the deep UV. 

While these relations have been extensively studied and will serve as explicit illustrations throughout this work, they should be understood as particular realizations within a broader class of MDRs. In particular, they exemplify qualitatively different UV behaviors, allowing one to assess in a controlled manner how inflationary observables respond to trans-Planckian modifications and to identify potential observational signatures in both the scalar and tensor sectors.

A key ingredient in the quantization of fields in time-dependent backgrounds is the choice of vacuum state. In cosmological spacetimes, a natural and widely adopted prescription is provided by adiabatic vacua. However, when including MDRs, the validity of the adiabatic approximation is no longer guaranteed a priori. This motivates a careful analysis of the conditions under which adiabaticity is preserved and adiabatic vacua remain well defined. 

A related question concerns the dependence of the quantization procedure on the choice of modes associated with different time variables (and the corresponding field rescaling). In general, different choices of time lead to distinct Fock representations. When MDRs are present, this issue becomes nontrivial and requires a careful assessment of the conditions under which such quantizations can still be regarded as physically equivalent.

Once a consistent notion of vacuum state has been established, a separate and equally essential issue concerns the definition of finite physical observables. A crucial aspect in any such analysis is the consistent removal of UV divergences from physical observables.  In particular, the energy-momentum tensor must be renormalized so that its expectation value remains finite and continues to serve as a reliable source in the semiclassical Einstein equations.  Without proper subtraction of divergent parts, one risks introducing spurious anomalies and violating energy conservation in the backreaction of quantum fields on the background geometry. Within this framework, the introduction of MDRs brings its own challenges, as the high‐momentum deformations alter the UV behavior and can generate new divergent structures.  In doing so, one must ensure that the extended subtraction preserves energy and momentum conservation, thereby maintaining the consistency of the semiclassical dynamics under Planck‐scale modifications.  In cosmological settings, adiabatic regularization~\cite{parker_toms_qft_curved_spacetime, BirrellDavies1982} is commonly employed to systematically remove UV divergences from composite observables, such as the two-point correlation function and, in appropriate cases, the trace of the energy-momentum tensor. This method has been extensively applied to the standard linear dispersion relation and to a variety of superluminal modified dispersion relations (see, e.g., \cite{parker_toms_qft_curved_spacetime, Nacir_2005}).

From the perspective adopted in this work, the applicability of adiabatic regularization is most transparently understood in terms of the UV asymptotic behavior of the frequency. As will be shown in the following sections, it is precisely this asymptotic structure that controls both the degree of divergence and the number of adiabatic subtraction terms required to render the theory finite. Within this general framework, the standard and Corley-Jacobson dispersion relations provide concrete and well-studied examples for which the adiabatic subtraction scheme can be implemented explicitly. The Unruh dispersion relation, although exhibiting a qualitatively different UV structure due to the saturation of the frequency at high momenta, can also be accommodated within the same general approach once its asymptotic behavior is properly taken into account. In this sense, these dispersion relations should be regarded as illustrative realizations of the broader class of asymptotically power-law MDRs considered here, rather than as isolated or exceptional cases.

This paper is organized as follows. In Sec.~\ref{Section II}, we present the key concepts of the canonical quantization of the field under consideration and formalize the notion of modified dispersion relations. We then review in Sec.~\ref{Section III} the concept of adiabaticity, extending the discussion made in previous studies. We derive the conditions for the validity of the adiabatic WKB expansion, ensuring that the zeroth-order adiabatic vacuum  remains well defined and that particle interpretation is meaningful even in the trans-Planckian regime. 
In Sec.~\ref{Section IV}, we consider a broad class of MDRs whose UV behavior is governed by a power law (or faster) and we analyze the conditions under which quantizations defined with different choices of time, together with a convenient time-dependent rescaling of the field variable, are unitarily equivalent. In Sec.~\ref{Section V}, we apply the framework of adiabatic regularization to renormalize the two-point correlation function for the broad class of MDRs considered before. Within this general setting, we illustrate the results by considering the standard dispersion relation, as well as the superluminal Corley-Jacobson and Unruh dispersion relations as representative examples.
Finally, in Sec.~\ref{Section VI}, we summarize our results and discuss the robustness of inflationary predictions in the presence of trans-Planckian modifications, together with a critical assessment of the limitations inherent to the renormalization scheme employed in this work.

\emph{Notation:} For the spacetime metric we use the signature \((-,+,+,+)\). We employ units such that \(c = \hbar = 1\). Furthermore, \(m_\textsc{p}\) denote the Planck mass.

\section{Preliminaries}
\label{Section II}
\subsection{Canonical quantization}
Consider a real, massive scalar field $\phi(x)$ minimally coupled to gravity propagating in spacetime.
The metric $g_{\mu\nu}$ will be treated as a given unquantized external field.
In the framework of classical field theory, 
the equations of evolution of the dynamical field $\phi$ take the Klein-Gordon~(KG) form
\begin{equation}
(\Box - m^2)\phi(x) = 0,
\label{TFM.KG}
\end{equation}
where $\Box$ is the D'Alembertian operator and $m$ is the mass of the field.

Given a globally hyperbolic spacetime, the KG inner product $(\phi_1,\phi_2)_\textsc{kg}$ is defined as in \cite{wald1984general}, where \( \phi_1 \) and \( \phi_2 \) belong to the space of complex solutions of Eq.~\eqref{TFM.KG}. A key property of the KG inner product is that it is preserved under temporal evolution and it is independent of the chosen spatial foliation of the spacetime.

We shall consider a Friedmann-Lemaître-Robertson-Walker
(FLRW) spacetime with flat space-like sections, described in the coordinate chart \((t,\vec{x})\), where \(t\) denotes cosmic time, \(\vec{x}\) the comoving spatial coordinates, and the line element takes the form
\begin{equation}
    \mbox{$ds^2=-dt^2+a^2(t)d\vec{x}^2$}. 
\end{equation}

Let \( \{ f_{\vec{k}}(t, \vec{x}), f_{\vec{k}}^*(t, \vec{x}) \} \) be a complete set of complex solutions of the KG equation, which we will refer to as mode functions, or simply modes. We require these modes to satisfy the orthonormality relations
\begin{align}
    (f_{\vec{k}}, f_{\vec{q}})_\textsc{kg} = \delta^{(3)}(\vec{k}-\vec{q}), 
    \qquad 
    (f_{\vec{k}}, f_{\vec{q}}^*)_\textsc{kg} = 0.
    \label{TFM.rel.ortonormalidad}
\end{align}

The most general real solution to the KG equation can then  be written as a linear combination of $f_{\vec{k}}$ and $f^*_{\vec{k}}$,
\begin{equation}
    \phi(t,\vec{x}) = \int d^3k \left(a_{\vec{k}} f_{\vec{k}}(t,\vec{x}) + a_{\vec{k}}^* f^*_{\vec{k}}(t,\vec{x})\right),
\end{equation}
where \( a_{\vec{k}} \) and \( a_{\vec{k}}^* \) are the so-called annihilation and creation coefficients, respectively.

Because of spatial homogeneity and isotropy, we can choose the mode functions $f_{\vec{k}}$ of the form
\begin{equation}
    f_{\vec{k}}(t, \vec{x})=\frac{1}{\sqrt{2(2\pi)^3a^3(t)}}F_k(t)e^{i\vec{k}\cdot\vec{x}},
    \label{TFM.modos f}
\end{equation}
with $k=|\vec{k}|$ and the functions $F_k$ satisfy the equation
\begin{equation}
    \ddot{F_k}+\omega_k^2F_k=0,
    \label{TFM. ecu.modos en t}
\end{equation}
where the dot is the derivative with respect to $t$ and
\begin{equation}
    \omega_k^2(t) \equiv \kappa^2(t)+m^2_\text{eff}(t),
\label{TFM.standar.dispersion.relation}
\end{equation}
where we have defined the time dependent effective mass 
\begin{equation}
       m^2_\text{eff}(t) \equiv m^2 -\frac{3}{2}\frac{\ddot{a}(t)}{a(t)}-\frac{3}{4}\frac{\dot{a}^2(t)}{a^2(t)}.
       \label{effective mass in t}
\end{equation}
Here $\kappa = k/a$ denotes the physical (observed) wavenumber, whereas \(k\) represents the comoving one.

In canonical quantization \cite{BirrellDavies1982}, the classical field is promoted to an operator by elevating the creation and annihilation coefficients to operators that satisfy standard commutation relations. The modes $f_{\vec{k}}(t, \vec{x})$ span the positive-norm subspace \(\mathbb{S}^+\) of the complexified space of solutions of the KG equation. The KG inner product restricted to \(\mathbb{S}^+\) is positive definite, and its completion defines the one-particle Hilbert space $\mathcal{H}$.
The Fock space associated with this choice of modes and time coordinate, denoted by \(\mathfrak{F} \), is then constructed as the symmetric Fock space built over \(\mathcal{H} \), $\mbox{$\mathfrak{F}  = \bigoplus_{n=0}^{\infty} \mathrm{Sym}\,\mathcal{H} ^{\otimes n}$}$.
Within this framework, the quantum field \(\hat{\phi}(x)\) is understood as an operator-valued distribution acting on \(\mathfrak{F}\), whose mode decomposition is determined by the choice of the positive-norm subspace \(\mathbb{S}^+\). Therefore, the field operator $\hat{\phi}$ on $\mathfrak{F}$ is given by
\begin{equation}
  \hat{\phi}(t, \vec{x}) = \int d^3k \left(f_{\vec{k}}(t, \vec{x}) \hat{a}_k + f_{\vec{k}}^*(t, \vec{x}) \hat{a}_k^\dagger\right).
\label{TFM.exp.mod.con f}  
\end{equation}

An alternative common method in the classical context to find solutions to the equation (\ref{TFM.KG}) is to express the FLRW metric in conformal time \(\eta\), related to cosmological time \(t\) by $dt=a(\eta)\,d\eta$. 
Analogously to the previous case, we write the modes in the following way
\begin{equation}
    \tilde f_{\vec{k}}(\eta,\vec{x}) = \frac{1}{\sqrt{2(2\pi)^3} a(\eta)} \tilde F_k(\eta) e^{i\vec{k}\cdot\vec{x}},
    \label{TFM.modos h}
\end{equation}
where the functions \( \tilde F_k \) satisfy (with \( ' \equiv \frac{d}{d\eta} \)):
\begin{equation}
    \tilde F_k'' + \tilde \omega_k^2 \tilde F_k = 0, \qquad \tilde \omega_k^2(\eta) \equiv a^2 \kappa^2(\eta)+ \tilde  m^2_\text{eff}(\eta),
    \label{TFM. eq. modos H}
\end{equation} 
with
\begin{equation}
    \tilde m^2_\text{eff}(\eta) \equiv m^2 a^2(\eta)- \frac{a''(\eta)}{a(\eta)}.
    \label{effective mass in eta}
\end{equation}
The expression for $\tilde \omega_k^2(\eta)$ is the standard dispersion relation in conformal time.
The choice of positive-norm modes in conformal time $\tilde f_{\vec{k}}(\eta,\vec{x})$ defines a corresponding positive-norm subspace \(\tilde{\mathbb{S}}^+\), a one-particle Hilbert space~\(\tilde{\mathcal{H}}\), and an associated Fock space \(\tilde{\mathfrak{F}}\). We will denote the field operator on \(\tilde{\mathfrak{F}}\) by  \(\hat{\tilde\phi}\).

More generally, we could carry out any reparametrization to any time variable $\sigma$  via the expression $d t=g(\sigma)d\sigma$, where $g$ is a positive monotonic non-vanishing function, and consider a convenient rescaling of the field modes so that their equation of motion is that of an oscillator without dissipative term and the corresponding quantization (which we will also denote by a tilde) obtained following the same procedure as above. This is done in appendix~\ref{app:a}.

In general, the Fock representations \(\mathfrak{F}\) and \(\tilde{\mathfrak{F}}\) need not be unitarily equivalent, a point that will be analyzed in detail in Sec.~\ref{Section IV}.

\subsection{Modified dispersion relations}

In standard quantum field theory in curved spacetimes, the dispersion relation for a free scalar field is given by Eq.~\eqref{TFM.standar.dispersion.relation} (or equivalently by Eq.~\eqref{TFM. eq. modos H} or \eqref{eq:DR generic time}). However, in high‐energy (or trans-Planckian) regimes—particularly when quantum‐gravitational effects are expected to become significant—this relation may require modifications~\cite{MartinBrandenberger2001}.

One convenient way to encode such trans‐Planckian corrections is to replace the usual dependence on the physical wavenumber squared $\kappa^2$ by a more general function \(\mathcal{K}(\kappa)\) in the expressions for the frequencies  \eqref{TFM.standar.dispersion.relation}, \eqref{TFM. eq. modos H}, and \eqref{eq:DR generic time}.
To modify the UV behavior, \(\mathcal{K}\) is endowed with a new characteristic momentum scale, \(\kappa_c\). For physical wavenumbers well below this threshold (\( \kappa \ll\kappa_c\)), one recovers the standard cuadratic relation  \(\mathcal{K} \approx  \kappa^2\), while for \( \kappa \gtrsim \kappa_c\) the function \(\mathcal{K}\) encodes the departure from relativistic behavior. Importantly, \(\kappa_c\) is introduced purely as a phenomenological scale and it merely marks the regime in which new high‐energy physics effects become relevant. Within an effective field theory framework, the function $\mathcal{K} $ can be constructed by expanding it as a power series in the physical momentum $\kappa$, with higher-order contributions suppressed by powers of the cutoff scale $\kappa_c$. Locality requires this expansion to involve only even powers of $\kappa$ \cite{corley1996}. Indeed, only local differential operators can appear in the effective action, which depend analytically on~$\kappa^2$. Odd powers would correspond to nonlocal, fractional operators and are therefore excluded.  

Multiple MDRs, characterized by different functional forms of $\mathcal{K}$, have been extensively studied in the literature. Among the most commonly discussed examples are the superluminal Corley-Jacobson (CJ) \cite{corley1996} dispersion relations and the Unruh \cite{unruh1995sonic} dispersion relation \cite{MartinBrandenberger2001, Niemeyer_2001}. These cases will serve here as illustrative realizations of a broader class of MDRs that are studied in this work.

The CJ dispersion relations,  originally discussed in~\cite{corley1996}, are obtained by truncating the expansion of $\mathcal{K}$ at fourth order. Depending on the sign of the $\kappa^4$ coefficient, one distinguishes between superluminal and subluminal modifications. Although both possibilities arise at this level, subluminal CJ relations are known to suffer from severe pathologies, including the breakdown of adiabaticity at high momenta, UV instabilities, and the absence of cutoff-independent predictions for the primordial power spectrum (see, e.g.,~\cite{MartinBrandenberger2001}). For this reason, when referring to this class of truncated models, we will restrict attention to the superluminal CJ dispersion relation,
\begin{equation}
   \mathcal{K} =\kappa^2+ {\kappa^4}/{\kappa_c^2},
   \label{TFM. K rara para CJ}
\end{equation}
where $\kappa = k/a$.

 As said before,  another widely studied example is provided by the Unruh dispersion relation \cite{unruh1995sonic},
\begin{equation}
  \mathcal{K}=\kappa_c^2\tanh^2\left( \kappa/\kappa_c\right),
  \label{kappa Unruh}
\end{equation}
which exhibits the characteristic feature that $\mathcal{K}$ saturates at the scale  $\kappa_c^2$  in the high-energy limit. Although a formal power-series expansion of $\mathcal{K} $ also yields a negative coefficient at order $\kappa^4$, the full dispersion relation remains strictly positive for all $\kappa$, thereby avoiding UV instabilities.

In this work, however, our analysis is not restricted to any specific functional form of $\mathcal{K}$. Instead, we consider a general class of MDRs whose UV behavior is governed by a power law (or faster UV growth). As will become clear throughout the paper, it is precisely this asymptotic behavior of the frequency that controls both the unitary equivalence between different quantizations and the renormalization properties of the theory.
From this perspective, the CJ and Unruh dispersion relations should be regarded as particular and physically illuminating examples of the general framework developed here. While they have been extensively studied in the literature, they do not possess features that are qualitatively distinct from those of the broader class of asymptotically power-law MDRs considered in this work.

\section{Adiabaticity}
\label{Section III}
\subsection{Adiabatic Vacua}
For simplicity, we restrict our analysis of this section to the coordinate chart $(t, \vec{x})$. The extension to conformal (or more general) time is straightforward and leads to equivalent results.

Defining a vacuum state in a time‐dependent background is nontrivial due to the absence of a global timelike Killing vector. In a FLRW spacetime, one seeks an approximation to the true ground state of the field when the background evolves slowly. This motivates the definition of an adiabatic vacuum: a state constructed so that each mode experiences only gradual changes in its effective frequency, minimizing particle production.

To implement this idea, one uses the Wentzel-Kramers-Brillouin (WKB) approximation to construct mode solutions to the field equation. We first note that \cite{BirrellDavies1982}
\begin{equation}
    F_{k}(t) \;=\; \frac{1}{\sqrt{W_{k}(t)}} \exp\!\Bigl(-i\!\int^{t} W_{k}(t')\,dt'\Bigr),
    \label{WKB_ansatz}
\end{equation}
is a solution of the mode equation \eqref{TFM. ecu.modos en t} provided that \(W_{k}(t)\) satisfies
\begin{equation}
    W_{k}^{2}(t) \;=\; \omega_{k}^{2}(t) \;+\; \left(\frac{3}{4}\,\frac{\dot{W}_{k}^{2}(t)}{W_{k}^{2}(t)} \;-\; \frac{1}{2}\,\frac{\ddot{W}_{k}(t)}{W_{k}(t)}\right).
    \label{W_equation}
\end{equation}

To solve this equation systematically, \(W_k(t)\) is expanded in an asymptotic (adiabatic) series, where each successive order involves higher time derivatives of the background quantities \cite{BenderOrszag1999}. Explicitly, one writes
\begin{equation}
    W_k(t) = \sum_{n=0}^{\infty} W_k^{(2n)}(t),
    \label{WKB_series}
\end{equation}
where \(W_k^{(2n)}(t)\) denotes the contribution of adiabatic order \(2n\).
All odd-order terms vanish identically. 

Substituting this series into Eq.~\eqref{W_equation} and matching order by order yields
\begin{align}
    W_{k}^{(0)}(t) &= \omega_{k}(t), \label{W0_def}\\
    W_{k}^{(2)}(t) &= \frac{3}{8}\,\frac{\dot{\omega}_{k}^{2}(t)}{\omega_{k}^{3}(t)} \;-\; \frac{1}{4}\,\frac{\ddot{\omega}_{k}(t)}{\omega_{k}^{2}(t)}. \label{W2_def}
\end{align}
Higher-order terms can also be computed, but we omit their expression since they are not particularly illuminating.

The adiabatic vacuum of order \(2n\) at an initial time \(t_{0}\) is defined by requiring that the exact mode functions and their first time derivatives coincide with their WKB approximations up to order \(2n\) at \(t=t_{0}\). Concretely, the zeroth‐order adiabatic vacuum is specified by the initial conditions
\begin{align}
    F_{k}(t_{0}) &= \frac{1}{\sqrt{2\,\omega_{k}(t_{0})}}, \nonumber\\
    \dot{F}_{k}(t_{0}) &= \left(-i\,\omega_{k}(t_{0}) \;-\; \frac{\dot{\omega}_{k}(t_{0})}{2\,\omega_{k}(t_{0})}\right)\frac{1}{\sqrt{2\,\omega_{k}(t_{0})}}.
    \label{zeroth_order_initial_conditions}
\end{align}
Higher‐order adiabatic vacua impose higher WKB order initial conditions for \(F_{k}\). 

Thus, while the adiabatic regime refers to the interval during which the frequency $\omega_k(t)$ varies slowly (justifying the WKB approximation), the adiabatic vacuum corresponds to the specific choice of initial conditions that align the mode functions with their WKB counterparts.

\subsection{Validity conditions}
Having introduced the notion of adiabatic vacua, we now turn to the question of when the adiabatic approximation is justified and what criteria govern its validity. If the background evolution significantly departs from adiabaticity, particle production can become substantial and may even dominate the energy density of the field. In the context of inflation, such backreaction could undermine the assumption that the inflaton potential is responsible for the accelerated expansion \cite{tanaka2000comment}. For these reasons, it is crucial to identify the conditions under which the WKB-based adiabatic approximation remains valid.

In the literature, two different criteria have been employed to ascertain the validity conditions of the adiabatic approximation. On the one hand, the  eikonal condition introduced in Ref.~\cite{Niemeyer_2001},
\begin{equation}
    \epsilon \equiv \left| \frac{\dot{\omega}_k}{\omega_k^2} \right| \ll 1,
    \label{TFM.eikonal.condition}
\end{equation}
is often taken as a sufficient indicator of adiabaticity. On the other hand, Ref.~\cite{Brandenberger2002} proposed the condition
\begin{equation}
\mathcal{Q} \equiv \left| \frac{3}{4} \left(\frac{\dot{\omega}_k(t)}{\omega_k^2(t)}\right)^2 
- \frac{1}{2} \frac{\ddot{\omega}_k(t)}{\omega_k^3(t)} \right| \ll 1,
\label{TFM.condiciones.necesarias.2}
\end{equation}
which is motivated by the consistency of the WKB expansion.

However, the actual requirement for the existence of the zeroth-order adiabatic vacuum is that the WKB frequency satisfies \(W_k(t) \simeq \omega_k(t)\). We find that ensuring this approximation demands more than imposing condition \eqref{TFM.condiciones.necesarias.2} alone. In fact, there exists an additional independent requirement,
\begin{equation}
\mathcal{I} \equiv \left| \int^{t} dt' \left(
\frac{3}{8} \frac{\dot{\omega}_k^2(t')}{\omega_k^3(t')}
- \frac{1}{4} \frac{\ddot{\omega}_k(t')}{\omega_k^2(t')}
\right) \right| \ll 1,
\label{TFM.condiciones.necesarias.1}
\end{equation}
which has so far remained unnoticed in the literature. This condition is inequivalent to both \eqref{TFM.eikonal.condition} and \eqref{TFM.condiciones.necesarias.2}, and is at least as restrictive as the latter.

Although the eikonal condition \eqref{TFM.eikonal.condition} is frequently assumed to be sufficient to guarantee adiabaticity in cosmological settings, we show that it does not, in general, imply either \eqref{TFM.condiciones.necesarias.2} or \eqref{TFM.condiciones.necesarias.1}. More precisely, we establish the relation
\begin{equation}
    \mathcal{Q} \ll 1 
    \quad \Longleftrightarrow \quad 
    \epsilon \ll 1 
    \ \text{and}\ 
    \left| \frac{\ddot{\omega}_k}{\omega_k^3} \right| \ll 1,
\end{equation}
which shows that the eikonal condition must be supplemented by an additional constraint in order to ensure~\eqref{TFM.condiciones.necesarias.2}.

This result clarifies the origin of the discrepancies observed in the numerical analysis of MDRs in de Sitter spacetime reported in Ref.~\cite{Macher_2008}. In that work, the eikonal condition \eqref{TFM.eikonal.condition} and the condition \eqref{TFM.condiciones.necesarias.2} are directly compared. However, as we have shown, the former does not imply the latter, and therefore cannot be regarded as a reliable indicator of adiabaticity by itself. Moreover, we emphasize that even condition \eqref{TFM.condiciones.necesarias.2} is not sufficient on its own: a proper assessment of adiabaticity requires that both necessary conditions \eqref{TFM.condiciones.necesarias.1} and \eqref{TFM.condiciones.necesarias.2} be simultaneously satisfied.

These conditions arise from truncating the WKB series, although this is not strictly conclusive: as an asymptotic expansion, the WKB series guarantees a vanishing relative error under suitable assumptions but does not control the absolute error. Nevertheless, their use is fully justified here, since a more rigorous yet practical analysis is not feasible within this framework \cite{Froman1996}.

A further subtlety in the analysis of adiabaticity concerns the validity of truncating the WKB expansion. Even when the adiabatic conditions derived above are satisfied, it is not guaranteed in full generality that the approximation \(W_k(t) \simeq \omega_k(t)\) remains valid. In particular, a pathological situation may arise if \(\omega_k(t)\) vanishes at some instant  while its first and second time derivatives also vanish at the same point, so that \(\omega_k\) exhibits a degenerate critical point at a zero. In such a case, the quantities entering the adiabatic conditions may remain small, while higher-order terms in the WKB expansion can become arbitrarily large compared to the leading-order contribution \(\omega_k\), thereby invalidating the adiabatic approximation.
To exclude this possibility, we shall henceforth restrict our analysis to frequencies \(\omega_k(t)\) that do not possess degenerate critical points at their zeros. This mild assumption ensures that the previously derived adiabatic conditions are robust and sufficient to guarantee \(W_k(t) \simeq \omega_k(t)\).

Independently of this general assumption, we have explicitly analyzed whether such degenerate zeros can occur in slow-roll (nearly de Sitter spacetime) inflationary background  for any MDR compatible with the standard infrared behavior \(\mathcal{K}(\kappa)\sim\kappa^2\) as \(\kappa\to0\). We find that this possibility is excluded: the simultaneous vanishing of \(\omega_k\) and its first two derivatives would require \(\mathcal{K}(\kappa)\) to be independent of \(\kappa\) at the relevant scale, in contradiction with the defining properties of physically admissible MDRs. Therefore, in the inflationary spacetime, the absence of degenerate critical points is guaranteed, and the adiabatic conditions derived above provide a fully reliable criterion for the validity of the adiabatic approximation.

Although in this section we have focused on the conditions ensuring the validity of the zeroth-order adiabatic approximation, higher-order adiabatic vacua can also be defined, as already mentioned. We do not derive general conditions for each adiabatic order here. Nevertheless, the procedure developed throughout this section can be systematically extended to obtain the corresponding conditions for the existence of higher-order adiabatic vacua. Such an extension leads to a hierarchical structure of increasingly restrictive requirements. In particular, the consistency of an adiabatic vacuum of a given order necessarily demands that all conditions associated with lower adiabatic orders be satisfied as well. This nested structure reflects the fact that higher-order adiabaticity builds upon the validity of the approximation at preceding orders.

\begin{center}
    \emph{A numerical example}
\end{center}

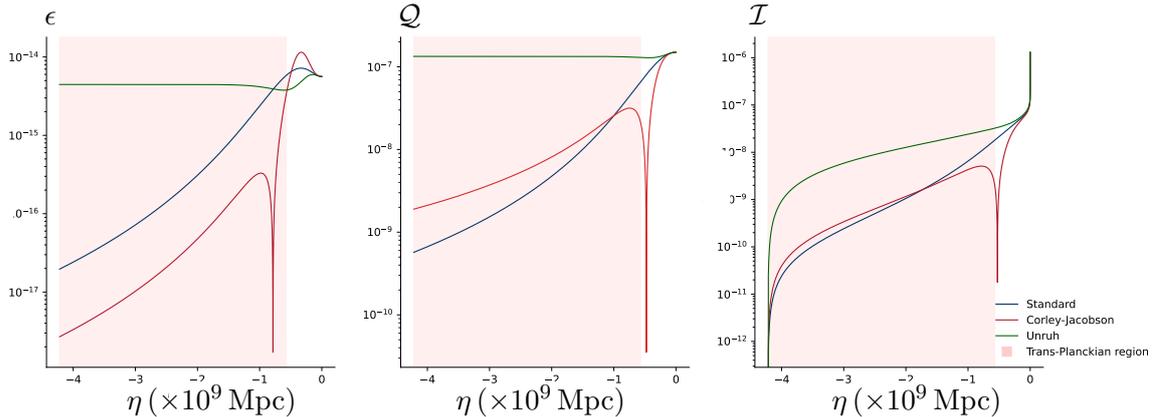
\begin{figure*}
  \centering
    \begin{overpic}[width=1\textwidth,keepaspectratio]{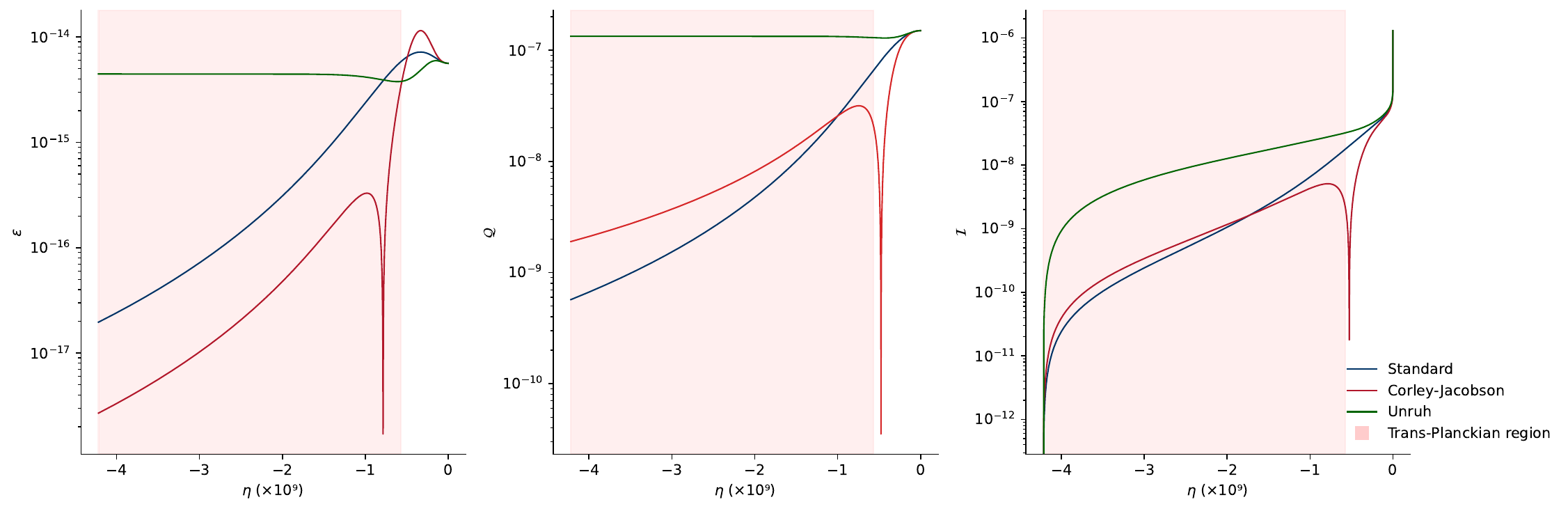}%
   \put(15,0){\color{white}\scalebox{4}[1.4]{$\blacksquare$}}
   \put(12,0){$\eta\,(\times10^{9}\,\mathrm{Mpc})$}
    \put(0,15){\color{white}\rotatebox{90}{\scalebox{4}[1.4]{$\blacksquare$}}}
     \put(5,33){{$\epsilon$}}

   \put(45,0){\color{white}\scalebox{4}[1.4]{$\blacksquare$}}
   \put(40,0){$\eta\,(\times10^{9}\,\mathrm{Mpc})$}
    \put(30,15){\color{white}\rotatebox{90}{\scalebox{4}[1.4]{$\blacksquare$}}}
     \put(35,33){{$\mathcal{Q}$}}

   \put(75,0){\color{white}\scalebox{4}[1.4]{$\blacksquare$}}
   \put(72,0){$\eta\,(\times10^{9}\,\mathrm{Mpc})$}
    \put(61,15){\color{white}\rotatebox{90}{\scalebox{4}[1.4]{$\blacksquare$}}}
     \put(65,33){{$\mathcal{I}$}}
     
    \end{overpic}

\caption{Joint evolution of $\epsilon$, $\mathcal{Q}$, and $\mathcal{I}$ as functions of the conformal time $\eta$ (in $\mathrm{Mpc}$), for the comoving mode $k = 10^{-2}\,\mathrm{Mpc}^{-1}$, which is the most infrared mode that remained trans-Planckian at the onset of inflation. The red-shaded region indicates the interval during which this mode was trans-Planckian.}
  \label{fig:tres_grandes}
\end{figure*}

Let us consider a model of cosmological inflation lasting approximately $60$ $e$-folds, and a scalar perturbation with mass
$m = 1.21 \times 10^{-6}\,m_\textsc{p}$ \cite{GarayGonzalezMartin2024}. In this inflationary scenario, we analyze the adiabaticity of the mode evolution by evaluating conditions \eqref{TFM.condiciones.necesarias.2} and \eqref{TFM.condiciones.necesarias.1}, and comparing them with the eikonal condition \eqref{TFM.eikonal.condition}. The corresponding numerical results are displayed in Fig.~\ref{fig:tres_grandes}.

We observe that, for all dispersion relations considered, each adiabaticity condition remains satisfied throughout the entire evolution. However, their numerical magnitudes exhibit a nontrivial hierarchy: in particular, the quantity $\epsilon$ is systematically much smaller than $\mathcal{Q}$.

As shown previously, there is no direct logical implication between these conditions. In many slowly varying backgrounds one may heuristically expect the second time derivative of the frequency to be suppressed with respect to the first, which could suggest that small values of $\mathcal{Q}$ naturally accompany small values of $\epsilon$. Nevertheless, this expectation is not guaranteed and does not hold in general.

The present numerical example explicitly illustrates this fact: although $\epsilon$ remains extremely small over the entire evolution, $\mathcal{Q}$ can take significantly larger values while still satisfying its own adiabatic bound. This demonstrates that the smallness of $\epsilon$ alone does not control the size of higher-derivative contributions, and therefore cannot be used as a reliable estimator of adiabaticity by itself.

\section{Comparison between quantizations}
\label{Section IV}

In the context of FLRW cosmology with modified dispersion relations, it is natural to ask whether the quantization performed in cosmic time is physically equivalent to the quantization carried out in conformal time. Since these choices lead to distinct mode decompositions, this issue must be addressed at the level of unitary equivalence between the corresponding Fock representations.
More generally, one may consider arbitrary time reparametrizations. The results of the analysis do not depend on the specific choice of time variable itself, but rather on the UV behavior of the associated mode frequencies at large values of the physical momentum $\kappa$, as we will see. Owing to this fact, and for the sake of simplicity, we restrict the explicit calculations in this section to quantizations defined with respect to cosmic and conformal time.

If there exists a unitary operator implementing the equivalence between the two quantizations, the corresponding mode functions must be related by a Bogoliubov transformation. In particular, the temporal dependence of the modes is required to satisfy
\begin{equation}
    F_k(t)=\alpha_k \tilde F_k(\eta(t))+\beta_k^* \tilde F_k^\dagger(\eta(t)).
\end{equation}
As a result, the problem of establishing unitary equivalence between both field quantizations  reduce to the standard criterion for Bogoliubov transformations in homogeneous and isotropic spacetimes. Namely, the Bogoliubov coefficient $\beta_k$ relating both sets of modes must be square integrable over momentum space, i.e.,
\begin{equation}
    \int_0^\infty  dk \, k^2\, \lvert \beta_k \rvert^2 < \infty.
    \label{beta condition}
\end{equation}

\subsection{Reduction to the UV behavior}

Making use of the WKB form for the mode functions \eqref{TFM.modos f} and \eqref{TFM.modos h}, the unitarity condition \eqref{beta condition} can be recast as
\begin{equation}
    \int_0^\infty dk \, k^2
    \frac{|G_k(\eta)|^2}
    {W_k(t(\eta)) \, \Tilde{W}_k(\eta)} < \infty,
    \label{final condition}
\end{equation}
where
\begin{equation}
    |G_k(\eta)|^2 \equiv \frac{1}{4}\Big( \frac{a'}{a}
     + \frac{W'_k(t(\eta))}{W_k(t(\eta))}
     - \frac{\Tilde{W}'_k(\eta)}{\Tilde{W}_k(\eta)} \Big)^2 +\Big(aW_k(t(\eta)- \Tilde{W}_k(\eta)\Big)^2.
    \label{crucial numerator}
\end{equation}

Since the WKB frequencies $W_k$ and $\tilde{W}_k$ are non-vanishing by construction, the corresponding mode functions are always well defined and properly normalized. In particular, $W_k$ does not vanish for any value of the physical momentum $\kappa$, including the infrared (IR) regime. This guarantees the absence of pathologies associated with ill-defined modes at small $\kappa$. Moreover, under these conditions, the integrand in \eqref{final condition} does not develop infrared divergences. While these two statements are closely related, it is important to distinguish between the non-vanishing of $W_k$ itself and the convergence of the integral in the IR, both of which are ensured in the present setting. 

As a consequence, any potential divergence of the integral \eqref{final condition} can only originate from the UV regime. The problem of unitary equivalence therefore reduces to the analysis of the asymptotic behavior of the integrand for large physical momenta $\kappa$.

Given the considerations above, it is sufficient to focus on MDRs that admit a well-defined asymptotic expansion in the UV. For such a broad class of MDRs, the WKB frequencies $W_k$ and $\tilde{W}_k$ are defined through an asymptotic expansion of the form \eqref{WKB_series}. As a result, the integral \eqref{final condition} cannot be evaluated exactly. However, an explicit computation is not required in order to determine its convergence properties.
Indeed, for MDRs that present an asymptotically power-law behavior, higher adiabatic orders exhibit an increasingly suppressed behavior in the UV. More precisely, each successive term in the WKB expansion contributes with an additional inverse power of the physical momentum $\kappa$.
Therefore, in order to assess the UV convergence of the integral \eqref{final condition}, it is sufficient to truncate the WKB expansions of both $W_k$ and $\tilde{W}_k$ at second adiabatic order. This truncation captures all potentially divergent contributions and considerably simplifies the analysis without loss of generality.

\subsection{UV behavior of MDRs}

We first consider a broad and physically well-motivated class of MDRs whose UV behavior is governed by a power law. Specifically, we assume that the frequency admits, in the UV regime, an asymptotic expansion of the form
\begin{equation}
    \omega_k \sim \frac{\kappa^{\alpha}}{\kappa_c^{\alpha-1}}
    \left[1 + \sum_{n \in \mathbb{Z}^+} c_n \,\left(\frac{\kappa_c}{\kappa}\right)^{n} \right],
    \label{asymptotic_dispersion}
\end{equation}
where $\alpha $ is a positive real number and the coefficients $c_n$ encode subleading corrections. The corresponding frequency $\tilde\omega_k$ associated with the alternative quantization exhibits the same leading UV behavior, up to time-dependent prefactors arising from the cosmological background.
Equation~\eqref{asymptotic_dispersion} is understood as an asymptotic expansion for $\kappa\to\infty$, rather than a convergent series. In particular, the difference between the exact dispersion relation and its asymptotic representation is subleading with respect to the leading term $\kappa^{2\alpha}$ and therefore contributes only at higher inverse powers of $\kappa$. Since the UV convergence of the Bogoliubov integral is entirely controlled by the leading asymptotic behavior, these subleading corrections do not affect the final result.

Polynomial dispersion relations constitute a particular case of this general framework, corresponding to integer values $\alpha>1$ and a finite number of subleading terms. Consequently, the analysis performed below applies equally to strictly polynomial MDRs as well as to more general dispersion relations admitting an asymptotic power-law expansion.

With these considerations in mind, we now determine the ultraviolet behavior of the integrand appearing in Eq.~\eqref{final condition}. Since our goal is to establish the leading scaling at large $\kappa$, we retain only the dominant UV contributions and systematically discard subleading terms.

Let us start with the numerator, $|G_k(\eta)|^2$. A careful and lengthy but straightforward asymptotic analysis of the combination shows that, in the large-$\kappa$ regime, 
\begin{equation}
\frac{a'}{a}+ \frac{W'_k}{W_k}- \frac{\tilde W'_k}{\tilde W_k}
\;\sim\;
\frac{\kappa^{-2\alpha}}{\kappa_c^{2-2\alpha}}
\left[1+\mathcal{O}\!\left(\frac{\kappa_c}{\kappa}\right)\right],
\label{termino W'-W' UV}
\end{equation} 
up to a multiplicative $\kappa$-independent (but time-dependent) factor.
Upon squaring this expression, the dominant behavior in the first term of $|G_k(\eta)|^2$ becomes $\kappa^{-4\alpha}$. Proceeding analogously, we analyze the second contribution to $|G_k(\eta)|^2$, namely the term involving the difference between the frequencies. A direct asymptotic expansion shows that, 
 \begin{equation}
\bigl(a W_k(\eta(t)) - \tilde W_k\bigr)^2
\;\sim\;
\frac{\kappa^{-2\alpha}}{\kappa_c^{2-2\alpha}}
\left[1+\mathcal{O}\!\left(\frac{\kappa_c}{\kappa}\right)\right].
\label{termno aW-W UV}
\end{equation}
Therefore, this second contribution to the numerator in Eq.~\eqref{final condition} scales as $\kappa^{-2\alpha}$ in the ultraviolet, with all subleading terms being suppressed by higher inverse powers of $\kappa$.

On the other hand, the denominator admits the asymptotic expansion
\begin{equation}
    \frac{1}{W_k \, \Tilde{W}_k}
    \;\sim\;
\frac{\kappa^{-2\alpha}}{\kappa_c^{2-2\alpha}}\left[1+
    \mathcal{O}\left(\frac{\kappa_c}{\kappa}\right)\right],
    \label{termino denominador}
\end{equation}
in the large-$\kappa$ limit.

Combining these results, the integrand in Eq.~\eqref{final condition} behaves asymptotically as
\begin{equation}
k^2
    \frac{|G_k|^2}
    {W_k \, \Tilde{W}_k} \sim \frac{\kappa^{2-4\alpha}}{\,\kappa_c^{4-4\alpha}}\left[1+
    \mathcal{O}\left(\frac{\kappa_c}{\kappa}\right)\right],
    \label{integrand in the UV}
\end{equation}
again, up to a multiplicative $\kappa$-independent factor.
Therefore, the integral converges if and only if $\alpha>3/4$.
Consequently, we conclude that all MDRs admitting an asymptotic power-law behavior with exponent $\alpha>3/4$ —including, as a particular case, polynomial dispersion relations— lead to unitarily equivalent quantizations in cosmological and conformal time.

Furthermore, it is straightforward to extend the above analysis to MDRs whose UV growth is faster than any power of $\kappa$. Indeed, one can prove that the integrand in~\eqref{final condition} scales as $\kappa^2/\omega_k^4$, so that it is enough that
\begin{equation}
    \omega_k \geq C \frac{\kappa^\alpha}{\kappa_c^{\alpha-1}},
    \qquad \text{for sufficiently large } \kappa,
    \label{faster}
\end{equation}
for some constant $C$ and $\alpha>3/4$, to guarantee convergence of the integral. We thus also conclude that MDRs exhibiting superpolynomial UV growth constitute a class for which the two quantizations under comparison are unitarily equivalent.

Moreover, had we chosen a different time variable 
$\sigma$ and quantized the corresponding rescaled field variable (see appendix \ref{app:a}), our conclusions would remain the same, since this different choice would only affect the time-dependent factors, but not the dependence on $\kappa$. In conclusion, as long as the considered MDR admits an asymptotic power-law behavior with exponent $\alpha>3/4$, or grows faster in the UV, the particular time variable chosen to describe the evolution of the field is irrelevant, provided that the quantized field variable verifies the equation of motion of a frequency-varying oscillator without dissipative term.

\subsection{Potentially divergent cases}

We now turn to the potentially divergent situations in which the UV behavior of the frequencies obeys a power law  characterized by an exponent $\alpha\leq3/4$ (including constant asymptotic behavior). In this case, the superficial degree of divergence of the integrand in Eq.~\eqref{final condition} is greater than~$-1$, and the corresponding integral diverges. The same applies for slower behavior.

The only possible mechanism to avoid this divergence would be for the Bogoliubov coefficients to vanish identically. This would require the condition for the comparison between cosmological and conformal time
\begin{equation}
    \beta_k = 0 \iff a W_k = \Tilde{W}_k
\end{equation}
to hold order by order in the WKB expansion, which is only true in the trivial case of Minkowski spacetime (in terms of the Minkowskian time). Although in this work we have not made the  relation between $W_k$ and $\tilde W_k$ explicit for a generic time reparametrization,
the criterion for unitary equivalence remains unambiguous: the associated Bogoliubov coefficient $\beta_k$ must vanish, which actually implies that the quantizations under comparison are identical.

\subsection{Standard, CJ, and Unruh dispersion relations}

In the literature, the cases of the standard dispersion relation \eqref{TFM. ecu.modos en t} and the superluminal dispersion relation \eqref{TFM. K rara para CJ} have been studied for different purposes. With regard to the unitary equivalence between quantizations in cosmological and conformal time, these two cases correspond to polynomial dispersion relations. Therefore, our previous analysis applies to them. Since both are characterized by $\alpha>3/4$, we conclude that the quantizations in different time coordinates are unitarily equivalent in both cases.

The Unruh dispersion relation constitutes a qualitatively distinct case. It falls into the category of divergent scenarios analyzed above.
Indeed, as follows from Eq.~\eqref{kappa Unruh}, the corresponding frequencies approach a constant in the UV regime for both quantizations. Consequently, all terms in the WKB expansion become independent of $\kappa$ in this limit, and the integrand in Eq.~\eqref{beta condition} tends to a constant. This leads to a quadratic UV divergence of the particle number density.
It is therefore sufficient to examine the zeroth-order contribution.

In the UV regime, the frequencies behave as
\begin{align}
    \omega^2_\textsc{uv} &= m_\text{eff}^2 + \kappa_c^2, \qquad
   \tilde \omega^2_\textsc{uv} = \Tilde{m}_\text{eff}^2 + a^2 \kappa_c^2.
   \label{omegas UV}
\end{align}
Since in non-trivial cosmologies these expressions do not satisfy $a \omega_\textsc{uv} = \tilde \omega_\textsc{uv}$, it follows that $\beta\neq 0$, and therefore the quantizations in cosmic and conformal time (or any other time variable, for that matter) are not unitarily equivalent for the Unruh dispersion relation, a fact that, to the best of our knowledge, has gone unnoticed in the previous literature.

\section{Adiabatic renormalization of the two-point correlation function}
\label{Section V}
Many physically meaningful quantities, such as the action and the energy-momentum tensor, are quadratic in the fields and their derivatives, evaluated at a single spacetime point. This local structure leads to well-known divergences in their expectation values. In flat spacetime, such divergences can often be handled by normal ordering in the case of free fields. However, the situation becomes more subtle in curved backgrounds: even free fields interact with the underlying geometry, giving rise to new types of divergences. Additionally, vacuum energy must be addressed with particular care, as it can have real gravitational consequences through its role in the Einstein field equations \cite{BirrellDavies1982, wald1994qft,parker_toms_qft_curved_spacetime}. The introduction of MDRs requires a reexamination of  the standard notion of the energy-momentum tensor, defined through the Hilbert prescription. Nevertheless the two-point function is still a key ingredient to construct it. 
Its relevance extends beyond its role in the computation of the energy-momentum tensor; it also holds intrinsic interest as its Fourier transform defines the power spectrum.

In this section, we present a general analysis of the adiabatic renormalization of the two-point correlation function for scalar fields with MDRs whose UV behavior is governed by a power law. Within this general framework, we explicitly discuss the standard dispersion relation, the superluminal CJ dispersion relation, and the Unruh dispersion relation as illustrative examples that have been extensively studied in the literature. These cases serve to exemplify the general renormalization procedure developed here, rather than constituting special or isolated scenarios. In particular, the CJ superluminal case is treated in detail for concreteness and for ease of comparison with previous results~\cite{Nacir_2005,LopezNacir:2008tx,LopezNacir:2009bhs,Rinaldi_2007}.

In the renormalization analysis that follows, it is necessary to specify the choice of quantization. For MDRs whose UV behavior is governed by an asymptotic power law with exponent $\alpha>3/4$, or faster UV growth, we have shown in the previous section that the quantizations associated with the mode functions $f_{\vec{k}}$ and $\tilde f_{\vec{k}}$ are unitarily equivalent. As a consequence, physical observables, as well as renormalized quantities, are independent of the particular time variable and the corresponding field redefinition adopted in the quantization procedure. 

Accordingly, and in order to facilitate a direct comparison with existing results in the literature, we shall work in what follows with the quantization defined in cosmological time $t$, namely that associated with the mode functions $f_{\vec{k}}(t,\vec{x})$.

For the Unruh dispersion relation, whose UV behavior is independent of $\kappa$, although the quantizations defined in cosmological and conformal times are not unitarily equivalent, we nevertheless adopt the quantization in cosmological time. As will be shown below, within the framework of adiabatic renormalization, performing the calculation in cosmological time or in conformal time leads to the same renormalized two-point correlation function, even when the corresponding quantizations are unitarily inequivalent. This does not represent a contradiction. Rather, it reflects the fact that adiabatic renormalization is a very specific subtraction scheme. By itself, the comparison of renormalized quantities—without further specification of the renormalization prescription and the physical criteria underlying it, which will be discussed later—is not sufficient to establish a clear or unambiguous comparison between different quantizations.

Let us consider a real scalar field minimally coupled to gravity, as introduced in Sec.~\ref{Section II}. 
The UV divergence in the two-point correlation function arises solely in the coincidence limit, that is, when the two spacetime points are brought together. In a homogeneous and isotropic background, this yields:
\begin{equation}
    \braket{\phi^2(t, \vec{x})}= \frac{1}{4\pi^2 a^3(t)} \int_0^\infty  \frac{dk \,k^2}{W_k(t)}=\frac{1}{4\pi^2} \int_0^\infty  \frac{d\kappa \,\kappa^2}{W_k(t)}.
    \label{TFM. 2-point. correl. func}
\end{equation}

This quantity generally presents divergences in the limit \( \kappa \to \infty \) which must be regularized and renormalized in order to obtain physically meaningful results.                                                   
The method of adiabatic subtraction proves to be particularly effective in cosmological settings~\cite{parker_toms_qft_curved_spacetime}. This procedure relies on the adiabatic expansion of mode functions, which provides a controlled approximation valid in the high-momentum regime. A key feature of this expansion is that each successive adiabatic order introduces additional inverse powers of the momentum \(\kappa\), leading to stronger suppression in the UV limit. Consequently, for dispersion relations in which the frequency \(\omega_k\) increases with~\(\kappa\), higher-order terms yield integrands that decay more rapidly at large momenta. This ensures that, beyond a certain order, the corresponding contributions to physical observables become UV finite. However, this property no longer holds if \(\omega_k\) ceases to depend on \(\kappa\), as is the case, for instance, with the Unruh dispersion relation. In such situations, the convergence of the adiabatic subtraction terms must be reassessed with care.

The scheme of adiabatic subtraction consists in removing, mode by mode, any adiabatic term that contains UV divergences for generic parameter values. Crucially, when a given adiabatic order contains any divergent contribution, the entire term must be subtracted—including its finite parts. By subtracting full adiabatic orders, one ensures the covariant conservation of the renormalized energy–momentum tensor \(\nabla_{\mu}\langle T^{\mu\nu}\rangle_{\rm ren} = 0\) in the case of the standard dispersion relation. When the divergent contributions depend on a parameter so that, for certain values, the integral becomes finite but, for others, it diverges, the subtraction must still be applied even if the raw integral converges. Enforcing this subtraction guarantees that the renormalized result remains a smooth, continuous function of the theory’s parameters, preserving both mathematical consistency and physical continuity \cite{parker_toms_qft_curved_spacetime}.

Let us now apply this renormalization scheme to Eq.~\eqref{TFM. 2-point. correl. func}. To this end, we expand the numerator in a series, formally obtaining
\begin{align}
  \frac{1}{W_k} &= \sum_{n} \left(W_k^{-1} \right)^{(2n)},
  \label{TFM.W-1}
\end{align}
where all odd-order terms vanish. Note that we have eliminated the $t$-dependence to alleviate the notation.

In particular, by using the expressions given in~\eqref{WKB_series}, we obtain:
\begin{align}
    \left(W_k^{-1} \right)^{(0)} &= \omega_k^{-1}, \nonumber\\
    \left(W_k^{-1} \right)^{(2)} &= \omega_k^{-2} W_k^{(2)}, \nonumber\\
    \left(W_k^{-1} \right)^{(4)} &= \omega_k^{-3} \left(W_k^{(2)}\right)^2 - \omega_k^{-2} W_k^{(4)}. \label{TFM.W-1^4}
\end{align}

 By substituting the results from Eq.~\eqref{TFM.W-1} and \eqref{TFM.W-1^4} in Eq.~\eqref{TFM. 2-point. correl. func}, one arrives at
 \begin{equation}
          \braket{\phi^2(t, \vec{x})} = \frac{1}{4\pi^2} \int_0^\infty d\kappa\, \kappa^2 \Big \{\left(W_k^{-1} \right)^{(0)} +\left(W_k^{-1} \right)^{(2)}+ \left(W_k^{-1} \right)^{(4)}+\ldots       \Big\},
     \label{TFM. Formula 2pcf expanded}
 \end{equation}
where the ellipsis denotes higher adiabatic orders not displayed.  This expression holds in full generality, irrespective of the particular dispersion relation employed.

In this framework, the adiabatic expansion plays a dual role: it makes the UV structure of the theory manifest, and it provides a systematic prescription for subtracting divergences in a manner that is both minimal and consistent. In doing so, it ensures that expectation values of composite observables—such as correlation functions and components of the energy-momentum tensor—remain finite and physically meaningful in curved spacetimes.

\subsection{Standard Dispersion Relation}
We begin by analyzing the standard dispersion relation. According to the general renormalization procedure, one must identify and subtract the terms in the adiabatic expansion of Eq.~\eqref{TFM. Formula 2pcf expanded} that lead to UV divergences.

In the high-momentum regime, the frequency asymptotically behaves as \(\omega_k \to \kappa\), so that the leading term in the integrand scales as \(\kappa^2\, \omega_k^{-1} \sim \kappa\). This contribution diverges quadratically and must therefore be removed. The next-to-leading-order term scales as \(\kappa^2 (W_k^{-1})^{(2)} \sim \kappa^0\), leading to a logarithmic divergence, which must also be subtracted. In contrast, the fourth-order term behaves as \(\kappa^2 (W_k^{-1})^{(4)} \sim \kappa^{-2}\), and is thus convergent in the UV regime.

Accordingly, the renormalized two-point function for the standard dispersion relation takes the form:
\begin{equation}
        \braket{\phi^2(t, \vec{x})}_\text{ren} = \frac{1}{4\pi^2 } \int_0^\infty d\kappa\, \kappa^2  \Bigg( 
    \frac{1}{W_k} - \left(W_k^{-1} \right)^{(0)} 
    - \left(W_k^{-1} \right)^{(2)} \Bigg).
    \label{TFM.2tcf standard ren.}
\end{equation}
which is a well-known result~\cite{parker_toms_qft_curved_spacetime}.

\subsection{Superluminal Dispersion Relation and power-law behaviors}

 Let us continue with the renormalization  of the two point function  for the superluminal CJ dispersion relation. We adopt the same procedure  as before. We recall that the dispersion relation was defined   in Eq.~\eqref{TFM. K rara para CJ}. Substituting into the general adiabatic expansion given in Eq.~\eqref{TFM. Formula 2pcf expanded}, one observes that in the UV limit $\omega_k \sim \kappa^2$, so that $\kappa^2\,\left(W_k^{-1} \right)^{(0)} \sim \kappa^0$ which produces a linear divergence.  Consequently, one must subtract this term to render the two-point function finite. For the next adiabatic correction, after a lengthly calculation, one finds that, in the UV regime, \(\kappa^2\bigl(W_k^{-1}\bigr)^{(2)} \sim \kappa^{-4}\), confirming that this higher-order term vanishes fast enough to yield convergence. Therefore, the two-point correlation function takes the form:
\begin{align}
    \langle \phi^2(t, \vec{x}) \rangle_{\mathrm{ren}} 
    &= \frac{1}{4\pi^2} \int_0^\infty \!\! d\kappa\, \kappa^2 
    \Bigl( \frac{1}{W_k} - \left(W_k^{-1} \right)^{(0)} \Bigr).
    \label{TFM.2tcf superluminal ren.}
\end{align}
This result was also found in \cite{Nacir_2005}.

It is worth noting that the CJ superluminal relation achieves stronger UV suppression than the standard case. Whereas the usual dispersion yields \(\omega_k\sim \kappa\) at large momentum (inducing quadratic divergences), the quartic correction enforces \(\omega_k\sim \kappa^2\), so that inverse‐frequency integrands decay more rapidly. This built‐in damping of high‐frequency modes illustrates how MDRs can serve as natural regulators for trans‐Planckian excitations, ensuring a well‐controlled quantum field behavior at ultra‐high energies.

The more general case of asymptotically power-law dispersion relations can also be renormalized using adiabatic subtraction. As shown above, the method consists of performing a WKB expansion in $1/W_k$ and subtracting a finite number of terms responsible for the UV divergences. From the analysis of the previous section, for a MDR of the form \eqref{asymptotic_dispersion}, the leading UV behavior of $1/W_k$ scales as $\kappa^{-\alpha}$, in agreement with \eqref{termino denominador}.  Throughout this discussion, we assume $\alpha>0$ (in fact, for superluminality, we need $\alpha>1$, although our analysis is also valid in the regime $0<\alpha\leq1$).  Otherwise, the following discussion in not directly applicable.

As discussed above, each successive adiabatic order in the WKB expansion introduces additional derivatives of the background quantities, which in the UV regime effectively lowers the power of $\kappa$ by $2\alpha$, provided $\alpha>0$. 
Under this assumption, the contribution to the two-point correlation function \eqref{TFM. 2-point. correl. func} coming from the adiabatic term $(W_k^{-1})^{(2n)}$ has a superficial degree of divergence \mbox{$\delta_n = 2 - (1+2n)\alpha$}. All the terms such that $\delta_{n}\geq -1$ will diverge and have to be subtracted in order to render the integral UV finite. 
This means that we have to subtract all the adiabatic terms $(W_k^{-1})^{(2n)}$, with $n\geq 0$ being an integer such that $2n\leq3/\alpha-1$.
Explicitly,
\begin{itemize}
    \item If $0<\alpha<1$, then we have to subtract terms with adiabatic order given by the general expression above.
    \item If $\alpha=1$, then we need to subtract the adiabatic terms $(W_k^{-1})^{(0)}$ and $(W_k^{-1})^{(2)}$.
    \item If $1<\alpha\leq3$, then we just need to subtract the lowest order adiabatic term $(W_k^{-1})^{(0)}$.
    \item If $3<\alpha$, the two-point function is already finite and there is no need for renormalization.
\end{itemize}
If we boldly extrapolate this analysis to $\alpha=0$, we directly conclude that all adiabatic terms diverge and have to be subtracted, giving a vanishing renormalized two-point function. This limiting case presents subtleties as we have mentioned and will be treated in detail in what follows for the particular case of the Unruh dispersion relation.
This prescription reflects the fact that, while the convergence condition is continuous in $\alpha$, the adiabatic subtraction scheme involves a discrete number of subtractions.

The standard dispersion relation ($\alpha=1$) and the superluminal CJ dispersion relation ($\alpha=2$) constitute two illustrative examples within the broader class of asymptotically power-law dispersion relations discussed above. In both cases, their UV behavior is characterized by a leading power-law term, which is precisely the feature that controls both the renormalization procedure and the unitary properties analyzed in this work. 

These results can be easily extended to the more general MDRs with  non-polynomic UV growths.

\subsection{Unruh Dispersion Relation}

While the superluminal dispersion relation ameliorates the UV behavior by accelerating the growth of the frequency with respect to the momentum, the Unruh dispersion relation exhibits a qualitatively different behavior: the frequency saturates at large momenta. This saturation introduces new obstacles in the renormalization procedure, which we now analyze in detail.

To examine this dispersion relation, recall the explicit form of Unruh’s frequency,
\begin{equation}
    \omega_k=\sqrt{\kappa_c^2\tanh^2\left({\kappa}/{\kappa_c}\right)+m_\text{eff}^2},
    \label{TFM.Unruh disp.rel}
\end{equation}
where we recall, $m_\text{eff}^2(t)$ is given by \eqref{effective mass in t} . One might then attempt to employ the adiabatic renormalization technique, successfully applied in the previous subsection, to renormalize the divergence of the two-point correlation function given by \eqref{TFM. Formula 2pcf expanded}. However, a crucial complication arises: the criterion governing the behavior of successive adiabatic orders in the WKB expansion can no longer be applied. Consequently, a more careful analysis of the UV limit of the series is required.

In Sec.~\ref{Section IV} we claimed that $W_k$ becomes independent of $k$ in the UV limit. We now provide a proof of this statement.
We analyze the divergences of the infinite series in \eqref{WKB_series}. Consider the generic structure of any term in the expansion \eqref{TFM.W-1}, which can be written as
\begin{equation}
    \left(W_k^{-1}\right)^{(2n)}
    =\sum_{m=1}^{N_{2n}} c_m^{(2n)} \,\omega_k^{-a_m} \prod_{l=1}^{M_m} \Bigl[\frac{d^{\,b_l}\omega_k}{dt^{\,b_l}}\Bigr]^{d_l},
    \label{TFM.General form W^-1}
\end{equation}
where $c_m^{(2n)}$ are real coefficients and $a_m, b_l, d_l, N_{2n}, M_m$ are positive integers. Each inverse adiabatic contribution is therefore a linear combination of powers of $\omega_k$ and its time derivatives.

In the UV limit, the $n$-th derivative of $\omega_k$ becomes independent of $\kappa$. Indeed, from Eq.~\eqref{TFM.Unruh disp.rel} we have $\tanh(\kappa/\kappa_c)\to1$ as $\kappa\to\infty$, so that all explicit $\kappa$-dependence disappears from these derivatives. Consequently, in the limit $\kappa\to\infty$, Eq.~\eqref{TFM.General form W^-1} reduces to
\begin{equation}
    \left(W_k^{-1}\right)^{(2n)} 
    \sim 
    \sum_{m=1}^{N_{2n}} c_m^{(2n)} \,\omega_\textsc{uv}^{-a_m - \sum_{l=1}^{M_m}d_l} 
    \prod_{l=1}^{M_m} 
    \Bigl[\frac{d^{\,b_l}m_\text{eff}^2}{dt^{\,b_l}}\Bigr]^{d_l},
    \label{TFM.General form W^-1 bis}
\end{equation}
where $\omega_\textsc{uv} $ was defined in Eq.\eqref{omegas UV}.

The absence of new divergences is not accidental; rather, it reflects the underlying physical mechanism. Effective frequencies that saturate in the UV naturally constrain the adiabatic expansion, ensuring that the renormalization procedure remains consistent with the fundamental structure of the theory.

Equation~\eqref{TFM.General form W^-1 bis} establishes the desired result: in the UV regime, the WKB expansion approaches a quantity that is independent of $\kappa$. As a consequence, the two-point correlation function exhibits a superficial degree of divergence $\delta_n=2$ at all orders in the asymptotic expansion.

We therefore define the renormalized two-point correlation function, following the adiabatic subtraction prescription, as the finite quantity obtained after removing all divergent contributions from the original expression, namely
\begin{equation}
    \braket{\phi^2(t,\vec{x})}_{\mathrm{ren}}= 0.
\end{equation}
Although this result may appear surprising, it should be noted that the two-point correlation function renormalized according to the Hadamard prescription also vanishes in Minkowski spacetime~\cite{wald1994qft}.

These results demonstrate that the Unruh dispersion relation is renormalizable order by order within the adiabatic expansion scheme. 
If the quantization is performed using conformal time, the corresponding frequency also saturates in the UV regime, and therefore the arguments presented above continue to apply. Although the explicit form of the leading term, which is independent of $\kappa$, differs from the one obtained in cosmological time, cf. Eq.~\eqref{TFM.General form W^-1 bis}, the structure of the WKB expansion remains qualitatively the same. In particular, all terms in the expansion exhibit the same degree of UV divergence and must therefore be subtracted. As a consequence, the renormalized two-point correlation function in conformal time also vanishes.

At first sight, this result may appear paradoxical, since the two quantizations under consideration are unitarily inequivalent. The resolution of this apparent contradiction lies in the fact that comparing renormalized quantities without further specification is, in general, not meaningful. A proper comparison of renormalized observables must also take into account the renormalized coupling constants arising from the counterterms that cancel the divergences. However, when adiabatic renormalization is implemented in a straightforward manner, one does not have explicit control over the individual counterterms. In particular, the choice of renormalization scheme implicitly fixes not only the divergent subtractions but also the finite remainders \cite{Collins:105730}.

As a result, the fact that two renormalized quantities coincide does not imply that the underlying quantizations are equivalent, nor does a discrepancy between them necessarily signal inequivalence. The comparison of renormalized observables, by itself, is therefore insufficient to draw conclusions about the unitary equivalence of different quantizations.

\section{Conclusions}
\label{Section VI}

In this work, we have analyzed the impact of MDRs on the quantization of fields in cosmological backgrounds, motivated by the trans-Planckian problem. Previous studies in the literature have focused on a limited set of specific examples, most notably the standard, superluminal CJ, and Unruh dispersion relations. 

We have shown, however, that many of the relevant physical properties associated with MDRs are largely controlled by their UV asymptotic behavior. This observation allows for a systematic generalization of earlier results to a broad class of dispersion relations whose high-energy regime is governed by an asymptotic power law (or faster). From this perspective, the dispersion relations most commonly analyzed in the literature should be regarded as particular realizations of the more general framework analyzed in this work.

The first part of our analysis shows that modifications of the UV behavior of the dispersion relation have a direct impact on the structure and validity of the adiabatic expansion used to define physically meaningful vacuum states. Within this framework, we have identified the conditions under which the WKB approximation, and consequently the zeroth-order adiabatic vacuum, can be consistently implemented.

It is important to emphasize, however, that these conditions should be understood within the intrinsic limitations of the adiabatic scheme itself. Since the WKB expansion is an asymptotic series that must be truncated at finite order, the resulting criteria cannot be regarded as strictly necessary and sufficient in a fully rigorous mathematical sense, but rather as reliable conditions within the domain of validity of the adiabatic approximation.

We have also carried out the analysis of unitary equivalence between Fock quantizations related by arbitrary time reparametrizations, together with a convenient time-dependent rescaling of the field variable. This rescaling brings the Fourier modes into the form of oscillators with time-dependent frequency and no dissipation, thereby rendering the canonical quantization procedure well defined. As shown in Appendix~\ref{app:a}, the question of unitary equivalence of these different descriptions can be addressed without loss of generality by comparing the quantizations defined in cosmological and conformal time, since smooth and monotonic time redefinitions (and the corresponding field rescaling) do not introduce new UV structures beyond those already encoded in the dispersion relation.

Within this general setting, we have studied quantizations associated with asymptotically power-law (or faster) dispersion relations and derived the precise condition under which unitary equivalence between different time choices is guaranteed. This analysis shows that unitary equivalence is controlled entirely by the leading UV behavior of the frequency.
Standard and superluminal CJ dispersion relations, which fall within this asymptotically power-law class, satisfy this condition and therefore lead to unitarily equivalent quantizations in cosmological and conformal time. By contrast, the Unruh dispersion relation violates this criterion, providing an explicit example in which quantizations associated with different time variables are not unitarily equivalent, and physical predictions may consequently depend on the chosen notion of time and quantized field variable.

A substantial part of this work has been devoted to the adiabatic renormalization of the two-point correlation function $\langle \phi^2 \rangle$ in the presence of modified dispersion relations. Within this general framework, we have derived explicit criteria that determine how adiabatic renormalization must be implemented, namely, which and how many adiabatic terms must be subtracted in order to render the theory UV finite. These criteria are entirely controlled by the leading asymptotic UV behavior of the dispersion relation.

We have shown that when the dispersion relation grows asymptotically only a finite number of adiabatic orders needs to be subtracted, and in some cases no subtraction is required at all. By contrast, when the leading UV behavior saturates to a constant value for large wave number, all adiabatic orders contribute equally to the UV divergences and must therefore be subtracted. 

The standard dispersion relation and the superluminal CJ modification, both extensively studied in the literature, provide representative examples of asymptotically power-law MDRs with a positive UV exponent, for which adiabatic renormalization involves the subtraction of only a finite number of terms. The Unruh dispersion relation, on the other hand, constitutes a paradigmatic example of the particular case with vanishing exponent, where the subtraction of all adiabatic orders is required.

These results suggest that, despite their phenomenological character, MDRs constitute a useful and internally consistent framework for probing the sensitivity of inflationary predictions to Planck-scale physics. At the same time, our analysis highlights the need for a careful control of the vacuum structure, the domain of validity of semiclassical approximations, and the renormalization scheme employed. These aspects are essential in order to extract physically meaningful conclusions. Further developments could extend the present analysis beyond de Sitter or slow-roll backgrounds, possibly within a more systematic effective field theory formulation.

Despite these encouraging results, several open issues remain. In particular, the connection between adiabatic renormalization and the counterterm-based renormalization schemes commonly employed in particle physics is not straightforward, although such a correspondence has been established for inflationary perturbations in the case of the standard dispersion relation~\cite{Markkanen_2018}. Moreover, within the adiabatic approach one does not retain direct control over the finite parts that are subtracted together with the divergences. As a result, the renormalized quantities obtained in this framework depend explicitly on the chosen renormalization scheme and should not be interpreted as scheme-independent physical predictions.
Combined with the absence of an explicit counterterm structure, this feature weakens the connection with the renormalization group and hinders a systematic, scheme-independent assessment of backreaction effects in generic MDR scenarios.

Building on the extensive analysis of the two-point function presented above and as a further remark on the implications of MDRs, let us briefly comment on the status of the energy-momentum tensor in the presence of MDRs. Although one may formally apply adiabatic subtraction to the quantity obtained from the Hilbert prescription, its interpretation as a physical source for Einstein’s equations becomes problematic once MDRs are introduced. The reason is that the matter action is no longer invariant under the full group of spacetime diffeomorphisms, and therefore the resulting object is not guaranteed to define a conserved current in the usual sense. While alternative constructions based on canonical Noether currents may be envisaged, their relation to semiclassical gravity is far from straightforward. A detailed and systematic analysis of these issues lies beyond the scope of the present work and will be addressed in future studies.

As a direction for future work that concerns both the two-point function and the energy and momentum, it would be desirable to establish a more direct link between the renormalization of cosmological observables and the standard framework of regularization and renormalization in quantum field theory, including its formulation in terms of the renormalization group. Such an approach could help to clarify the scheme dependence of the results and place the treatment of backreaction effects in MDR models on firmer theoretical grounds.

\section*{Acknowledgments}
We are grateful to Gabriel Álvarez Galindo for his invaluable assistance in the more rigorous and formal discussions concerning the interpretation and mathematical subtleties of the WKB series and asymptotic expansions.
This work was financially supported by Grant PID2023-149018NB-C44 funded by
MCIN/AEI/10.13039/501100011033 and by ``ERDF/EU''.
RN was funded by the Royal Society through the University Research Fellowship Renewal URF$\backslash$R$\backslash$221005. CDR acknowledges financial support from
MCIU (Ministerio de Ciencia, Innovación y Universidades, Spain) fellowship FPU24/03217.

\appendix
\section{Arbitrary time reparametrizations}
\label{app:a}

In the main text, we have discussed two different quantization procedures: one associated with the mode functions $f_k(t,\vec{x})$ defined with respect to cosmological time $t$, and another one associated with the modes $\tilde f_k(\eta,\vec{x})$ defined in conformal time $\eta$, related through $dt = a(\eta)\, d\eta$, with their respective mode rescaling. We now generalize this construction by considering an arbitrary time variable $\sigma$.

Let $\sigma$ be related to cosmological time through a positive, monotonic function $g(\sigma)$ according to $ dt = g(\sigma)\, d\sigma $. Using this reparametrization in Eq.~\eqref{TFM. ecu.modos en t}, the mode equations can be written as those of a time-dependent oscillator. As discussed in Sec.~\ref{Section II}, the presence of a dissipative term can be eliminated by an appropriate field rescaling. In the present case, this is achieved by defining
\begin{equation}
    f_k(t(\sigma),\vec{x}) = \sqrt{\frac{g}{a^3}}\, \tilde f_k(\sigma,\vec{x}) .
\end{equation}
With this choice, the equation of motion for the temporal part $\tilde F_k(\sigma)$ of the rescaled modes $h_k(\sigma,\vec{x})$ takes the form
\begin{equation}
    \tilde F_k'' + \Big( g^2 \kappa^2 + M^2_{\text{eff}} \Big) \tilde F_k = 0 ,
    \label{ecu modos g y sigma}
\end{equation}
where the effective mass squared is given by

\begin{equation}
      M^2_{\text{eff}} = g^2 m^2 
    + \frac{1}{2} \left( \frac{g''}{g} - 3 \frac{a''}{a} - \frac{g'^2}{g^2} + 3 \frac{a'^2}{a^2} \right) - \frac{1}{4} \left( 3 \frac{a'}{a} - \frac{g'}{g} \right)^2 .\label{eq:DR generic time}
\end{equation}

Notice that choosing $g=1$ reproduces the mode equation in cosmological time, Eq.~\eqref{TFM. ecu.modos en t}, while choosing $g=a$ yields the corresponding equation in conformal time, Eq.~\eqref{TFM. eq. modos H}.

From Eq.~\eqref{ecu modos g y sigma}, it is clear that the dependence on the specific choice of time reparametrization is entirely encoded in the effective mass term $M^2_{\text{eff}}(\sigma)$. The dependence on the physical momentum $\kappa$ is unaffected, apart from an overall factor of $g^2$. Therefore, smooth and monotonic time reparametrizations do not introduce any new UV structure in the mode equations, and unitary equivalence or inequivalence between quantizations is entirely determined by the asymptotic behavior of the dispersion relation rather than by the particular choice of time variable.

Consequently, the analyses presented in Secs.~\ref{Section IV} and \ref{Section V} can be straightforwardly extended to arbitrary time variables and associated field-rescaling by following the same steps, with the only modification being the time-dependent factors associated with the effective frequency.

\bibliographystyle{apsrev4-2}
\bibliography{biblio}

@article{Unruh1995sonic,
  author    = {W. G. Unruh},
  title     = {Sonic analogue of black holes and the effects of high frequencies on black hole evaporation},
  journal   = {Phys. Rev. D},
  volume    = {51},
  pages     = {2827},
  year      = {1995},
  doi       = {10.1103/PhysRevD.51.2827}
}

@inbook{MartinBrandenberger2001,
author = {JÉRÔME Martin and ROBERT H. Brandenberger},
title = {A COSMOLOGICAL WINDOW ON TRANS-PLANCKIAN PHYSICS},
booktitle = {The Ninth Marcel Grossmann Meeting},
chapter = {},
doi = {10.1142/9789812777386_0456},
URL = {https://www.worldscientific.com/doi/abs/10.1142/9789812777386_0456}
}

@article{Brandenberger2001,
  author    = {R. H. Brandenberger and J. Martin},
  title     = "{The robustness of inflation to changes in super-Planck-scale physics}",
  journal   = {Mod. Phys. Lett. A},
  volume    = {16},
  pages     = {999},
  year      = {2001},
  doi       = {10.1142/S0217732301004170}
}

@article{Brandenberger2002,
  author    = {R. H. Brandenberger and J. Martin},
  title     = {On signatures of short distance physics in the cosmic microwave background},
  journal   = {Int. J. Mod. Phys. A},
  volume    = {17},
  number    = {25},
  pages     = {3663--3680},
  year      = {2002},
  doi       = {10.1142/S0217751X02010765}
}

@article{Brandenberger_2013,
   title={Trans-Planckian issues for inflationary cosmology},
   volume={30},
   ISSN={1361-6382},
   url={http://dx.doi.org/10.1088/0264-9381/30/11/113001},
   DOI={10.1088/0264-9381/30/11/113001},
   number={11},
   journal={Class. Quantum Grav.},
   publisher={IOP Publishing},
   author={Brandenberger, Robert H and Martin, Jérôme},
   year={2013},
   month=apr, pages={113001} }

@article{Niemeyer_2001,
    author = "Niemeyer, Jens C. and Parentani, Renaud",
    title = "{Transplanckian dispersion and scale invariance of inflationary perturbations}",
    doi = "10.1103/PhysRevD.64.101301",
    journal = "Phys. Rev. D",
    volume = "64",
    pages = "101301",
    year = "2001"
}

@book{BirrellDavies1982,
  author    = {N. D. Birrell and P. C. W. Davies},
  title     = {Quantum Fields in Curved Space},
  publisher = {Cambridge University Press},
doi={https://doi.org/10.1017/CBO9780511622632},
  year      = {1982},
  isbn      = {9780521278584}
}

@book{wald1984general,
  author    = {Wald, Robert M.},
  title     = {General Relativity},
  publisher = {University of Chicago Press},
  year      = {1984},
  doi       = {10.7208/chicago/9780226870373.001.0001},
  url       = {https://doi.org/10.7208/chicago/9780226870373.001.0001}
}

@online{tanaka2000comment,
  author = {Tanaka, T.},
  title = "{A comment on trans-Planckian physics in inflationary universe}",
  eprint = {astro-ph/0012431},
  archivePrefix = {arXiv},
  primaryClass = {astro-ph},
  year = {2000}
}

@book{Froman1996,
  author    = {N. Fröman and P. O. Fröman},
  title     = {Phase-Integral Method: Allowing Nearlying Transition Points},
  publisher = {Springer-Verlag},
  year      = {1996},
  doi       = {https://doi.org/10.1007/978-1-4612-2342-9}
}

@article{GarayGonzalezMartin2024,
  author    = {L. J. Garay and M. L. González and M. Martín-Benito and R. B. Neves},
  title     = "{Adiabatic approach to the trans-Planckian problem in loop quantum cosmology}",
  journal   = {Phys. Rev. D},
  volume    = {109},
  pages     = {123534},
  year      = {2024},
  doi       = {10.1103/PhysRevD.109.123534}
}

@book{kolbturner,
  author    = {Kolb, Edward W. and Turner, Michael S.},
  title     = {The Early Universe},
  publisher = {Westview Press},
  year      = {1994},
  isbn      = {9780201626742},
  url       = {https://www.adlibris.com/nb/bok/the-early-universe-9780201626742}
}

@article{wmap2011,
  author    = {{G. Hinshaw et al.}},
  title     = {Nine-Year Wilkinson Microwave Anisotropy Probe (WMAP) Observations: Cosmological Parameter Results},
  journal   = {Astrophys. J. Suppl.},
  volume    = {208},
  pages     = {19},
  year      = {2013},
  doi       = {10.1088/0067-0049/208/2/19}
}

@article{corley1996,
  author    = {S. Corley and T. Jacobson},
  title     = {Black hole lasers},
  journal   = {Phys. Rev. D},
  volume    = {54},
  pages     = {1568},
  year      = {1996},
  doi       = {10.1103/PhysRevD.54.1568}
}

@article{PhysRevD.63.123501,
  title = {Trans-Planckian problem of inflationary cosmology},
  author = {Martin, J\'er\^ome and Brandenberger, Robert H.},
  journal = {Phys. Rev. D},
  volume = {63},
  issue = {12},
  pages = {123501},
  numpages = {16},
  year = {2001},
  month = {May},
  publisher = {American Physical Society},
  doi = {10.1103/PhysRevD.63.123501},
}

@article{Macher_2008,
  author       = {Macher, Jean and Parentani, Renaud},
  title        = "{Signatures of trans-Planckian dispersion in inflationary spectra}",
  journal      = {Phys. Rev. D},
  volume       = {78},
DOI={10.1103/physrevd.78.043522},
  year         = {2008},
  pages        = {043522},
  archivePrefix = {arXiv},
  primaryClass = {gr-qc}
}

@book{parker_toms_qft_curved_spacetime,
  author    = {Leonard E. Parker and David J. Toms},
  title     = {Quantum Field Theory in Curved Spacetime: Quantized Fields and Gravity},
  year      = {2009},
  publisher = {Cambridge University Press},
  isbn      = {978-0-521-87787-9},
  url       = {https://www.cambridge.org/9780521877879},
}

@book{mukhanov2005physical,
  title        = {Physical Foundations of Cosmology},
  author       = {Mukhanov, Viatcheslav},
  year         = {2005},
  publisher    = {Cambridge University Press},
  isbn         = {9780521563987},
  doi          = {10.1017/CBO9780511790553}
}

@book{wald1994qft,
  author    = {Wald, Robert M.},
  title     = {Quantum Field Theory in Curved Spacetime and Black Hole Thermodynamics},
  publisher = {University of Chicago Press},
  year      = {1994},
  isbn      = {9780226870274},
  url       = {https://press.uchicago.edu/ucp/books/book/chicago/Q/bo3684008.html}
}

@article{Markkanen_2018,
doi = {10.1088/1475-7516/2018/05/001},
url = {https://doi.org/10.1088/1475-7516/2018/05/001},
year = {2018},
month = {may},
publisher = {},
volume = {2018},
number = {05},
pages = {001},
author = {Markkanen, Tommi},
title = {Renormalization of the inflationary perturbations revisited},
journal = {JCAP}
}

@article{Nacir_2005,
    author = "Lopez Nacir, D. and Mazzitelli, F. D. and Simeone, C.",
    title = "{Renormalized stress tensor for trans-Planckian cosmology}",
    doi = "10.1103/PhysRevD.72.124013",
    journal = "Phys. Rev. D",
    volume = "72",
    pages = "124013",
    year = "2005"
}

@book{Collins:105730,
  author    = {John C. Collins},
  title     = {Renormalization},
  publisher = {Cambridge University Press},
  year      = {1984},
  doi       = {10.1017/CBO9780511622656},
  isbn      = {9780511622656, 0521311772},
}

@misc{achúcarro2022inflationtheoryobservations,
      title={Inflation: Theory and Observations}, 
      author={Ana Achúcarro et. al},
      year={2022},
      eprint={2203.08128},
      archivePrefix={arXiv},
      primaryClass={astro-ph.CO},
      url={https://arxiv.org/abs/2203.08128}, 
}

@book{BenderOrszag1999,
  author    = {Carl M. Bender and Steven A. Orszag},
  title     = {Advanced Mathematical Methods for Scientists and Engineers: Asymptotic Methods and Perturbation Theory},
  publisher = {Springer New York},
  year      = {1999},
  isbn      = {9780387989310},
  doi       = {10.1007/978-1-4757-3069-2},
}

@article{LopezNacir:2009bhs,
    author = "Lopez Nacir, D. and Mazzitelli, Francisco D.",
    editor = "Bezerra, V. B. and Mostepanenko, V. M. and Romero, Carlos",
    title = "{On the renormalization procedure for quantum fields with modified dispersion relation in curved spacetimes}",
    doi = "10.1142/S0217751X09045005",
    journal = "Int. J. Mod. Phys. A",
    volume = "24",
    pages = "1565--1569",
    year = "2009"
}

@article{LopezNacir:2008tx,
    author = "Lopez Nacir, D. and Mazzitelli, F. D.",
    title = "{Renormalization in theories with modified dispersion relations: Weak gravitational fields}",
    doi = "10.1016/j.physletb.2009.01.025",
    journal = "Phys. Lett. B",
    volume = "672",
    pages = "294--298",
    year = "2009"
}

@article{Rinaldi_2007,
    author = "Rinaldi, Massimiliano",
    title = "{Momentum-space representation of Green's functions with modified dispersion on ultrastatic space-time}",
    journal = "Phys. Rev. D",
    volume = "76",
    pages = "104027",
    year = "2007",
    doi = "10.1103/PhysRevD.76.104027",
    publisher = "American Physical Society (APS)"
}
\end{document}